\documentstyle[12pt,bezier]{article}
\def \sect #1 {\setcounter{equation} 0\section{#1}}
\def \be  {\begin{equation}}
\def \ee  {\end{equation}}
\def \ba  {\begin{eqnarray}}
\def \ea  {\end{eqnarray}}
\def \baa {\begin{eqnarray*}}
\def \eaa {\end{eqnarray*}}
\def \bb  {\begin{thebibliography}}
\def \eb  {\end{thebibliography}}
\newcommand \ci [1] {\cite{#1}}
\newcommand \bi [1] {\bibitem{#1}}
\def \lab #1 {\label{#1}}
\newcommand\re[1]{(\ref{#1})}
\def \qqquad {\qquad\quad}
\def \qqqquad {\qquad\qquad}
\newcommand\lr[1]{{\left({#1}\right)}}

\def \vev #1 {\langle{#1}\rangle}
\def \VEV #1 {\left\langle{#1}\right\rangle}
\newcommand \ket [1] {|{#1}\rangle}
\newcommand \bra [1] {\langle {#1}|}
\def \e {\mbox{e}}
\def \CO {{\cal O}}

\def \B {{\rm B}}
\def \b {{\rm b}}
\def \s {{\rm s}}
\def \u {{\rm u}}
\def \barnu {{\bar\nu}}
\def \xm {{M_B}/{M}}

\newcommand\fr[2]{\lr{\frac{#1}{#2}}}
\def \fracs #1#2 {\mbox{\small $\frac{#1}{#2}$}}
\def \partder #1 {{\partial \over\partial #1}}
\def \bin #1#2 {{\left({#1}\atop{#2}\right)}}
\def \as {\relax\ifmmode\alpha_s\else{$\alpha_s${ }}\fi}
\def \alpi {\frac \as \pi}
\def \al #1 {\frac {\as({#1})}{\pi} }
\def \ds #1 {\ooalign{$\hfil/\hfil$\crcr$#1$}}
\def \MS {\overline{MS}}

\def \E   {{}_{{}_E}}
\def \GeV {\mbox{GeV}}

\begin{document}
\def\thefootnote{\fnsymbol{footnote}}
\thispagestyle{empty}
\hfill\parbox{35mm}{{\sc ITP--SB--94--35} \par
                         hep-ph/9407344   \par
                         July, 1994}
\vspace*{45mm}
\begin{center}
{\LARGE Infrared factorization in inclusive B meson decays}
\par\vspace*{22mm}\par
{\large Gregory~P.~Korchemsky}%
\footnote{On leave from the Laboratory of Theoretical Physics,
          JINR, Dubna, Russia}
and
{\large George Sterman}
\par\bigskip\par\medskip
{\em Institute for Theoretical Physics, \par
State University of New York at Stony Brook, \par
Stony Brook, New York 11794 -- 3840}
\end{center}
\vspace*{20mm}
\begin{abstract}

We perform infrared factorization of differential rates of radiative and
semileptonic inclusive decays of the $\B$ meson in the end-point
region of photon and charged lepton spectrum, respectively, in the
leading heavy quark mass limit. We find that the differential rates are
expressed in terms of hard, jet and soft functions, which satisfy evolution
equations. Solving these equations, we find expressions for the moments
of the differential rates in their end-point regions, which take into
account all leading and nonleading logarithmic corrections in perturbation
theory, as well as large nonperturbative power corrections.
Expanded to the one-loop level, our predictions coincide with the results
of existing lowest order calculations for $\B\to\gamma X_s$ and
$\B\to l\barnu X_u$. Nonperturbative corrections appear in our formalism
from the boundary value of the soft function in the evolution equation.
The soft function is a universal process-independent function, which
describes the distribution of the $\b$ quark in the $\B$ meson.
Its behavior in the end-point region is governed by the nonperturbative
asymptotics of a Wilson line expectation value. By considering the
contributions of infrared renormalons, we find an ansatz for the Wilson
line, which leads to a Gaussian model for the heavy quark distribution
function.
\end{abstract}

\newpage
\def\thefootnote{\arabic{footnote}}
\setcounter{footnote} 0

\section{Introduction}

Radiative and semileptonic decays of $\B$ mesons near maximum photon and
charged lepton energy, respectively, are of special interest for the
determination of electroweak parameters and the detection of new physics.
They also present special theoretical challenges \ci{review} because of
large perturbative and nonperturbative corrections from QCD.

Recently, Bigi {\it et al.}\ci{Bigi} and Neubert \ci{Neub} analyzed
nonperturbative corrections to these
processes in the $\Lambda/M$ expansion, with $M$ the heavy quark
mass and $\Lambda$ the usual QCD scale.  They found that the leading
term in this expansion is described by the same universal function,
$f(x)$, for both processes.
This function is nonperturbative in origin, and describes the $\B$ meson
decay outside the phase space available for the decay of a free $\b$ quark.
They also
related the first few moments of $f(x)$ to hadronic matrix elements
in heavy quark effective theory (HQET).  As pointed out in
\ci{ACM,Bigi,Wise}, however, perturbative corrections are also large
in the end-point region $x=E/E_{max}\sim 1$ and need to be resumed.
It was observed that the leading
(double-logarithmic) corrections, $(\alpha_s\ln^2(1-x))^k$,
exponentiate \ci{ACM}.
Clearly, a complete picture of inclusive
$\B$ decay in the end-point region requires a unified treatment of
perturbative and nonperturbative contributions.  Our goal in
this paper is to develop such a formalism, based on well-known
factorization theorems and evolution equations in perturbative
QCD \ci{theorems}, and on the analysis of infrared renormalons
\ci{IRR1}--\ci{IRR3} in the
evolution of Wilson lines, or path-ordered exponentials.

In Section 2, we apply the factorization technique
\ci{tech,S,YaF} to inclusive
radiative decay $\B\to\gamma X_s$ in the photon end-point region.
The relevant factorization is valid to leading power in $1/M$, and
involves two functions, 
one of which is the
distribution $f(x)$ describing
soft gluon emission by the $\b$ quark, and the other a ``jet" function $J(x)$,
describing collinear interactions of the outgoing $\s$ quark.
The ``universality" of $f(x)$ is a direct
consequence of factorization. In Sections 3 and 4 we show
the close relation of the distribution and jet functions to
Wilson lines. We then derive the evolution of these functions,
which resums both leading and nonleading logarithms as $x\to 1$.
These equations are analogous to the usual GLAP \ci{GLAP}
evolution equations, and follow directly from the
renormalization properties of Wilson lines and quark fields.
The solutions to these equations give an
expression for $d \Gamma_{\B\to \gamma X_s}/d x$ that
includes all large perturbative corrections near the
end-point.  As in most evolution equations, they require a
boundary value, which summarizes nonperturbative corrections.
In Section 5, we show that the perturbative
solution contains infrared renormalons, which introduce
ambiguities in perturbation theory.  These ambiguities occur
precisely in the powers $\Lambda^2/[(1-x)M]^2$ that we expect from
nonperturbative effects. This analysis lead us to
a nonperturbative estimate for the relevant Wilson lines, and
as a consequence, to a Gaussian model for the nonperturbative
behavior of the
distribution function.  Finally, we extend our analysis to
the end-point region in semileptonic decays, and show that the
only significant difference is that these decays require an
extra integral over the phase space of the neutrino.

\section{Factorization in radiative decay of B-meson}

Let us consider the inclusive radiative decay $\B\to\gamma X_s$ in the limit
when the mass of the $\s$ quark is neglected. In the rest frame of the $\B$
meson we define a scaling variable $x$ as the ratio of the photon energy
to the mass of the $\b$ quark,
$$
x=\frac{2E_\gamma}{M} \,.
$$
The condition $M_X^2>0$ for the final state $X_s$ leads to the restriction
\ci{Bigi,Neub},
$$
0 < x < {M_B}/{M} \sim 1.09 \,,
$$
where $M_B$ is the mass of the $\B$ meson.
We are interested in the inclusive distribution, $\frac{d\Gamma}{dx}$,
with the energy of photon near the end-point, $x\sim 1$.
The decay $\B\to\gamma X_s$ occurs through the transition $\b\to \s\gamma$,
which is described by an effective hamiltonian \ci{Effe}. In the heavy
quark limit, $M\to\infty$, the momentum of the $\b$ quark is
$
Q_\mu = M v_\mu\,,
$
where $v_\mu$ is the velocity of the $\B$ meson. As was stressed in
\ci{Bigi,Neub},
perturbation theory describes a free $\b$ quark decay in the
region $x < 1$, and in order to penetrate inside the ``window''
$1 < x < M_B/M$
one has to take into account nonperturbative effects.
In the rest frame of the $\B$ meson, the momenta of the photon and $\b$ quark,
$q_\mu$ and $Q_\mu$, respectively, may be taken to have the following
light-cone components
$(q_\pm=\frac1{\sqrt 2}(q_0\pm q_3)$ and $\vec q_t=(q_1,q_2))$
\be
q_\mu\equiv(q_+,q_-,\vec q_t)=\frac{M}{\sqrt 2}(x,0,\vec 0)\,,\qquad
Q_\mu=\frac{M}{\sqrt 2}(1,1,\vec 0)\,.
\lab{q}
\ee
Then, the momenta of the $\s$ quark, $P=Q-q$, is given by
$$
P_\mu=\frac{M}{\sqrt 2}(1-x,1,\vec 0)\,.
$$
In the end-point region, $x\to 1$, the $\s$ quark moves in the minus direction
with a high energy $\sim M/2$ and has a small invariant mass, $P^2=M^2(1-x) \ll
M^2$. Thus, for $x\sim 1$, the $\s$ quark can produce a jet of collinear
particles accompanied by soft radiation,
from which we expect large perturbative and nonperturbative corrections.
This is exactly the same situation
one encounters analyzing the behaviour of hadronic processes (DIS structure
functions, Drell-Yan cross section) in perturbative QCD near the boundary of
phase space \ci{tech}.
That is why the resummation technique developed in \ci{tech,S,YaF}
applies to the inclusive $\B$ decay in the end-point region.
In particular, analyzing diagrams of fig.1 contributing to the decay
$\B\to \gamma X_s$ we find the following three configurations of particle
momenta associated with leading $1/M$ behavior
(analog of leading twist in DIS):
$$
\begin{array}{ll}
\mbox{Hard} (H): &  k_+ \sim k_- \sim k_t  = \CO(M) \,,
\\
\mbox{Jet} (J):  &
k_+ = \CO(M(1-x))\,,\quad k_- =\CO(M)\,, \quad k_t = \CO(M\sqrt{1-x})\,,
\\
\mbox{Soft} (S): &  k_+ \sim k_- \sim k_t = \CO(M(1-x))\,.
\end{array}
$$
For the $x\to 1$ limit, the hard subprocess gets contributions only from
virtual particles. Notice that the minus and transverse components of the
momenta of particles in the jet subprocess are much larger than those in
the soft subprocess. That is, the jet subprocess carries almost all
the $P_-$ and $P_t$ momentum, while the small $P_+$ momentum is distributed
between the jet and soft
subprocesses. In individual diagrams of fig.1, particles from jet and soft
subprocess interact with each other. In the sum of all
diagrams in the leading $1/M$ limit, however,
the contribution of hard, jet and soft
subprocesses may be factorized into the form \ci{theorems,tech}
\be
\frac1{\Gamma_\gamma}\frac{d\Gamma}{dx} = \frac{M}{v_+}
\int_{(l_+)_{min}}^{P_+} dl_+\ S(l_+) J(P_+ - l_+) H(P_-)
\,,\qquad
P_+=\frac{M}{\sqrt 2}(1-x)\,.
\lab{fac}
\ee
Here,
$\Gamma_\gamma=\Gamma(b\to s\gamma)
=\frac{\alpha G_F^2}{32\pi^4}M^5 |V_{tb}V^*_{ts}|^2 C_7^2(M)$
is the partonic total width in the
Born approximation \ci{Jezabek},
and $l_+$ is the light-cone component
of the total momentum of soft gluons emitted in the partonic subprocess.
Let us make an important comment about the integration limits for $l_+$ in
\re{fac}. The upper limit $(l_+)_{max}=P_+$ follows from the condition that
the momentum of the $\s$ quark jet, $P-l$, be time-like.
The lower limit $(l_+)_{min}$
corresponds to the minimal energy of soft gluons in the final state.
In perturbation theory soft gluons are emitted by $\b$ and $\s$
quarks into the
final state and momentum conservation requires
$(l_+)_{min}=0$. In a realistic $\B$ meson, however,
soft gluons may also take energy from the light components of the $\B$ meson.
Thus, the minimal energy of soft gluons emitted by the $\b$ and $\s$ quarks in
the partonic subprocess may even be negative nonperturbatively,
\be
(l_+)_{min}=-(M_B-M)/\sqrt{2}\,,
\lab{lmin}
\ee
which leads a window for $1<x<M_B/M$.

To separate the subprocesses $H$, $J$ and $S$ in momentum space
one has to introduce \ci{theorems} a factorization scale $\mu$. The
contribution of each subprocess depends on this scale, although the
$\mu-$dependence cancels in the differential rate \re{fac}. In particular,
the contribution of the hard subprocess, $H$, depends only on $M$ and
$\mu$.\footnote{In what follows the dependence of hard, soft and jet
subprocesses on the coupling constant $\as(\mu)$ is implied.}

\section{The soft subprocess}

In emitting soft gluons, the $\b$ and $\s$ quarks behave as classical
relativistic particles. That is, all effects of their interactions with soft
gluons are factorized into eikonal phases given by the path-ordered
exponentials \ci{eik=WL,eik}, or Wilson lines, $P\exp(i\int_C dz\cdot A(z))$,
evaluated along their classical trajectories $C$ with the gauge field
$A(z)$ describing the soft radiation. As a consequence, the contribution of the
soft subprocess to the differential rate is given by a Fourier transformed
expectation value of a Wilson line%
\footnote{Notice that the gauge fields are ordered along the integration
          path $C$ and not according to time. However, on different parts
          of $C$ path-ordering implies time or anti-time ordering \ci{Wn}.}
\ci{S,Wn}
\be
S(l_+)=Mv_+\int_{-\infty}^{\infty}\frac{dy_-}{2\pi} \e^{i y_-l_+}
W_C((v\cdot y)\mu)\,,
\qquad
W_C\equiv \bra{B} P\exp\lr{i\int_C dz\cdot A(z)} \ket{B}\,,
\lab{S}
\ee
where the integration over $y_-$ fixes the total momentum of soft gluons in
the final state to be equal to $l_+$ and $\ket{B}$ denotes the $\B$ meson
state. The integration path $C$ in the definition of $W_C$ is shown
in fig.2. It goes from $-\infty$
to point $0$ along the quark velocity $v$, then along the light-cone minus
direction to point $y_-$, and then from $y_-$ to $-\infty$ along $-v$.

The following comments are in order. Notice that according to the
definition, $S(l_+)$ depends only on the properties of the bound state
$\ket{B}$ and not on the particular short distance subprocess. This is why
$S(l_+)$ is a universal distribution. In fact, the
function $S(l_+)$ was previously defined in \ci{S} in the analysis of large
$x$ behavior of the structure function of DIS.
It can be easily checked that the heavy quark distribution function
\ci{Bigi,Neub} coincides with the definition of $S(l_+)$.
Indeed, in the leading $1/M$ limit, heavy quark fields are equivalent to
path-ordered exponentials \ci{KR}, and making this identification
one gets $S(l_+)=F(x)=f(k_+)$ with $l_+=-\bar\Lambda x/\sqrt 2=-k_+/\sqrt 2$
in the notation of refs.\ci{Bigi} and \ci{Neub}, respectively.

In perturbation theory, the Wilson line $W_C$ obeys the
following relations \ci{S,Wn},
$$
 W_C=W(y_-v_+\mu-i\varepsilon)
=W^*(y_-v_+\mu-i\varepsilon)
=W(-y_-v_+\mu+i\varepsilon)\,,
$$
which ensures the reality of $S(l_+)$. Here, $\mu$ is the factorization
scale and the
``$-i\varepsilon$'' prescription comes from the analogous property of the
free gluon propagator and is thus of
perturbative origin.  Let us define a dimensionless variable $z$ and rewrite
\re{S} as
\be
S(l_+)\equiv f(z,M/\mu)
=\int_{-\infty}^{\infty}\frac{d\sigma}{2\pi} \e^{i\sigma(1-z)}
W(\mu\sigma/M - i\varepsilon)\,,\qquad
l_+=\frac{M}{\sqrt 2}(1-z)\,.
\lab{f}
\ee
where $\sigma=y_-v_+M$.
Then, the ``$-i\varepsilon$'' prescription immediately leads to the spectral
property \ci{S}
$$
f(z,M/\mu)=0,  \qquad \mbox{for $z > 1$}\,,
$$
which implies that there is no ``window'', $z>1$, in perturbation theory.
For this window to appear one has to take into account nonperturbative
corrections to the Wilson line $W_C$ in \re{S}.

The heavy quark distribution function defined above, $f(z,M/\mu)$, gets large
perturbative corrections $\sim \lr{\frac{\log^n(1-z)}{1-z}}_+$ from the region
$z\to 1$ corresponding to $l_+\to 0$ in \re{fac}.  Analyzing the moments of
$f(z,M/\mu)$ in perturbation theory we find the following remarkable relation
\ci{Wn}
\be
f_n(M/\mu)\equiv \int_0^{\xm} dz z^{n-1} f(z,M/\mu)
=W(-in\mu/M) + \CO(1/n) \,,
\lab{fn}
\ee
which means that the large $n$ behavior of the moments of the heavy quark
distribution function is given by the Wilson line expectation value
$W_C$ evaluated along the path $C$ with the formal identification
$y_-v_+=-in/M$.
This suggests that in the large $n$ limit one has to treat $W_C$ as a
nonlocal functional of the gauge field \ci{S,Wn}, rather than to expand it
into a divergent power series in $y_-$ using the operator product expansion
\ci{Bigi,Neub}.

We now use the renormalization properties \ci{ren,2-loop,WL}
of the Wilson line $W_C$
as a function of $\mu$, which depend on the particular form of the path $C$.
The integration path $C$ of fig.2 has two cusps at points $0$ and $y$ and a
light-like segment in between. As a consequence, the light-like Wilson
line $W_C$ obeys the renormalization group equation \ci{S,Wn}
\be
{\cal D}\ W_C=
-\left\{\Gamma_{cusp}(\as)
           \left[\log(\rho-i\varepsilon)+\log(-\rho+i\varepsilon)\right]
        +\Gamma(\as)
 \right\} W_C,\qquad
\rho=y_-v_+\mu\ \e^{\gamma_{\E}}\,,
\lab{RG}
\ee
where ${\cal D}\equiv \mu\frac{\partial}{\partial\mu}
+\beta(g)\frac{\partial}{\partial g}$, $\gamma_{\E}$ is the Euler
constant and
the anomalous dimensions $\Gamma_{cusp}(\as)$ and $\Gamma(\as)$ are known
to two-loop order \ci{2-loop,Wn}.
Solving this equation, and using the relation \re{f},
we find for the moments of the distribution function
\be
f_n(M/\mu)=\exp\lr{
\int_\mu^{Mn_0/n}\frac{dk_t}{k_t}
\left[2\Gamma_{cusp}(\as(k_t))\log\frac{k_tn}{Mn_0}
                                  +\Gamma(\as(k_t))\right]
}f_n^{(0)}\,,
\lab{soln}
\ee
where $n_0=\e^{-\gamma_{\E}}$ and
$f_n^{(0)}$ 
has a nonperturbative origin, as the boundary value of the
Wilson line in the solution of the
RG equation \re{RG}. From this expression, we find that the moments obey the
following evolution equation \ci{S,Wn},
\be
{\cal D}\ f_n(M/\mu)=-\left(2\Gamma_{cusp}(\as)\log\frac{\mu n}{Mn_0}
                                  +\Gamma(\as)\right)f_n(M/\mu)\,,
\lab{eq1}
\ee
where the r.h.s. depends on the normalization point through
$\log\frac{n\mu}{n_0M}$.

\section{The jet subprocess}

The jet subprocess $J$ in eq.\re{fac} describes the decay of the $\s$ quark
with momentum
$P-l$\/ into a jet of collinear particles. The invariant mass of the jet is
$2P_-(P_+-l_+)=M^2(z-x)\geq 0 $, and the contribution of the jet
subprocess depends only on this quantity and on the renormalization scale
$\mu$,
\be
J(P_+-l_+)\equiv J(2P_-(P_+-l_+),\mu^2)=J(M^2(z-x),\mu^2)\,.
\lab{J}
\ee
To take into account large perturbative corrections from the
region $z\to x$ we again consider the moments:
\be
J_n(M/\mu)\equiv M^2 \int_0^1 dz z^{n-1} J(M^2(1-z),\mu^2)\,.
\lab{Jn}
\ee
After integration of the r.h.s., using the relation $z^n\approx\e^{-n(1-z)}$
in the large $n$ limit, the function $J_n$ depends only on two scales, $M^2/n$
and $\mu^2$, which allows us to write
\be
J_n(M/\mu)=J_{n\mu^2/M^2}(1)+\CO(1/n)\,.
\lab{fun}
\ee

Let us now substitute expressions \re{f} and \re{J} into \re{fac},
using \re{lmin},
\be
\frac1{\Gamma_\gamma}\frac{d\Gamma}{dx}=
\int_x^{\xm} dz \ f(z,M/\mu)\ M^2 J(M^2(z-x),\mu^2)\ H(M/\mu)\,,
\lab{new}
\ee
and consider the moments of the differential rate defined as
\be
{\cal M}_n\lr{\B\to \gamma X_s}\equiv \frac1{\Gamma_\gamma}
\int_0^{\xm} dx\ x^{n-1}\ \frac{d\Gamma}{dx} \,.
\lab{Mn}
\ee
The behavior of $d\Gamma/dx$ in the end-point region $x\sim 1$
corresponds to the large $n$ asymptotics of the moments
${\cal M}_n\lr{\B\to \gamma X_s}$.
We notice that for $z,\, x\sim 1$ one can replace $J(M^2(z-x),\mu^2)\approx
J(M^2(1-x/z),\mu^2)$ in \re{new}. Then, the moments
of the differential rate factorize \ci{tech} into the product of moments of
the distribution function $f_n$ and the moments of the jet $J_n$
defined in \re{fn} and \re{Jn},
\be
{\cal M}_n\lr{\B\to \gamma X_s}
=
 f_n(M/\mu) J_n(M/\mu)  H(M/\mu) + \CO(1/n)
\lab{res}
\ee
The condition that the l.h.s. of this relation does not depend on the
renormalization point $\mu$ leads to the following RG equations,
\ba
{\cal D}\ J_n(M/\mu)&=&\left(2\Gamma_{cusp}(\as)\log\frac{\mu^2n}{M^2n_0}
                                  -2\gamma(\as)\right)J_n(M/\mu)\,,
\lab{eq2}
\\
{\cal D}\ H(M/\mu)&=&\left(-2\Gamma_{cusp}(\as)\log\frac{\mu}{M}
                                  +2\gamma(\as)+\Gamma(\as)\right)H(M/\mu)\,,
\nonumber
\ea
where the functional form of $J_n$, eq.\re{fun}, and
independence of $H$ on $n$ were used.
The anomalous dimensions entering into these equations have the following
one-loop expressions \ci{S,Wn}
\be
\Gamma_{cusp}(\as)=\alpi C_F\,,\qquad
\Gamma(\as)=-\alpi C_F\,,\qquad
\gamma(\as)=-\frac34\alpi C_F \,.
\lab{one}
\ee
Here, $\gamma(\as)$ coincides with the one-loop quark anomalous
dimension in the axial gauge. To understand this property we notice
that the collinear subprocess $J$ can be defined as a cut propagator of
the $\s$ quark in the light-like axial gauge $(n\cdot A)=0$ with
$n_\mu=(0,y_-,\vec 0)$. Indeed, in this gauge the interaction between
soft gluons and particles from the jet subprocess is supressed in
each diagram of fig.1 and the factorization \re{fac} is manifest.
Then, the evolution equation \re{eq2} describes the renormalization
properties of a cut quark propagator in the light-like axial gauge.

Solving the RG equations \re{soln} and \re{eq2} for $S_n$, $J_n$ and $H$,
we find after some algebra our complete expression for the moments of the
decay distribution,
\ba
{\cal M}_n\lr{\B\to \gamma X_s}&=&f_n^{(0)}\ C_\gamma(\as(M))
\nonumber
\\
&&\hspace{-25mm}
\times
\exp\left(
-\int_{n_0/n}^1 dy
\left[
2 \int_{My}^{M\sqrt{y}} \frac{dk_t}{k_t}
 \Gamma_{cusp}\lr{\as(k_t)}
+\Gamma\lr{\as(My)}
+\gamma\lr{\as(M\sqrt y)}
\right]
\right)
\lab{res1}
\ea
where $C_\gamma=1+\CO(\as(M))$ takes into account $\log^0(n)$
corrections to ${\cal M}_n$.

\section{Infrared renormalons in Wilson lines}

For large $n$, the integrals in \re{res1} get dominant contribution
from the region $y\sim 1/n$, and for $n\sim M/\Lambda$ we encounter
the singularities of the coupling constant. This means that perturbation
theory becomes ill-defined in the end-point region of the photon spectrum,
and we have to take into
account nonperturbative corrections to $d\Gamma_{\B\to \gamma X_s}/dx$.
As discussed in \ci{IRR1}, both perturbative
and nonperturbative contributions to physical quantities are ambiguous
at the level of power corrections and only their sum is unique \ci{IRR1,KS}.
This allows us to understand the structure of nonperturbative effects
by analyzing the ambiguity of the perturbation series associated with
the so-called infrared renormalons.

In the end-point region we expect large nonperturbative corrections
to the Wilson line entering the definition \re{S} of the distribution
function. Let us examine the contribution of IR renormalons to $W_C$
by performing an ``improved'' calculation of $W_C$ in perturbation
theory. Using the definition \re{S} we find the one-loop contribution
to $W_C$ \ci{Wn} and use the nonabelian exponentiation theorem for path
ordered exponentials \ci{Gath} to get
\be
W_C=\exp\lr{
            C_F\mu^{4-D}\int\frac{d^Dk}{(2\pi)^{D-2}}\ \as\delta_+(k^2)
            \lr{\frac{y}{(yk)}-\frac{v}{(vk)}}^2
            \lr{1-\e^{i(yk)}}\lr{1-\e^{-i(yk)}}
            }\,,
\lab{WC1}
\ee
where $y=(0_+,y_-,\vec 0)$, $d^Dk=dk_+dk_-d^{D-2}\vec k$, and where
$\mu$ is the scale parameter of the dimensional
regularization with $D=4-2\varepsilon$. The choice of the argument of the
coupling constant $\as$ is determined by
higher order corrections to $W_C$ to be the
transverse momentum of the gluons \ci{eik,kt}.
Now let us substitute $\as=\as(\vec k^2)=\int_0^\infty d\sigma
\exp(-\sigma\beta_0\log\frac{\vec k^2}{\Lambda^2})$ into \re{WC1} and perform
the integration over gluon momenta to get
\be
W_C=\exp\lr{C_F \frac{(4\pi\mu^2/\Lambda^2)^\varepsilon}
                         {\Gamma(1-\varepsilon)}
\int_{\varepsilon/\beta_0}^\infty
d\sigma\ \lr{\Lambda y_-v_+}^{2\sigma\beta_0}
\Gamma(-2\sigma\beta_0)\cot(\pi\sigma\beta_0)
(1-\sigma\beta_0)
}\,.
\lab{WC2}
\ee
Because $(\Lambda y_-v_+)^{2\sigma\beta_0}=\exp(-\sigma/\as(1/y_-v_+))$,
the exponent has a form of an inverse Borel transformation
\ci{IRR1}--\ci{IRR3}. Integration over
small Borel parameter $\sigma$ corresponds to large transverse momenta
$\vec k^2$ and gives rise to ultraviolet poles in $\varepsilon$.  After
their subtraction in the $\MS$ scheme, one verifies that $W_C$ satisfies
the RG equation \re{RG}. We now recognize that away from
$\sigma=\varepsilon/\beta_0$ the integrand in \re{WC2} has infrared
renormalon singularities generated by the $\Gamma-$function at points
$\sigma^*=\frac1{2\beta_0},\frac1{\beta_0},\frac3{2\beta_0},
...$. Thus, to give a meaning to the perturbative
expansion one has to fix a prescription for integrating these
singularities \ci{IRR3}. Different choices of the prescription lead to
results which differ in power corrections of the form
$(\Lambda y_-)^{2\sigma^*\beta_0}$. The leading power corrections
correspond to the left-most singularity. Note that the
singularity of the $\Gamma-$function at $\sigma^*=\frac1{2\beta_0}$ is
compensated by the cotangent, so that the leading IR renormalon appears
at $\sigma^*=\frac1{\beta_0}$. Thus, resummed perturbation theory
generates (but fails to describe uniquely)
$\CO(\Lambda^2 y_-^2)$ power corrections to the Wilson line $W_C$.
At the same time, this means that nonperturbative effects should also
contribute to $\CO(\Lambda^2 y_-^2)$ power corrections to make the Wilson line
$$
W_C=W_{pert.}(y_- ) W_{nonpert.}(y_- )
$$
well defined. In summary, the exponentiation of
perturbative corrections to $W_{pert.}(y_-)$
implies that IR renormalons appear in the
exponent of $W_{pert.}(y_-)$. Hence, in order to compensate their contribution
one can choose the ``minimal'' ansatz for nonperturbative part
$W_{nonpert.}(y_-)$:
\be
W_{nonpert.}(y_-)=\exp\lr{- (y_- v_+)^2/l^2 + \CO(y_-^3)}\,,
\lab{ans}
\ee
where $l$ is a nonperturbative correlation length. Notice that the
expansion in the exponent does not contain a term linear in $y_-$.

Let us test the ansatz of eq.\re{ans}
by performing an expansion of the Wilson line $W_C$ defined by
\re{S} in powers of $y_-$. We denote a straight Wilson line
as $\Phi_v(x)=P\exp(i\int_{-\infty}^0 ds v\cdot A(vs+x))$ and apply
the identity $P\exp(i\int_0^1 ds y_- A_+(ys))=\exp(-y_- D_+)
\exp(y_-\partial_+)$
to get
$$
W_C=\bra{B} \Phi_v^\dagger(y_-) P\e^{i\int_0^1 ds y_- A_+(ys)} \Phi_v(0)
\ket{B}
   =1+\sum_{k=1}^\infty \frac {(iy_-)^k}{k!}
\bra{B} \Phi_v^\dagger(0) \ \lr{iD_+}^k\ \Phi_v(0) \ket{B}
$$
where $D_\mu=\partial_\mu-iA_\mu$
is the covariant derivative in the quark representation.
According to its definition, the Wilson line satisfies the relation
$(v\cdot D)\Phi_v(x)=0$, which is similar to the equation of motion of
the heavy quarks in HQET \ci{Geor}. This allows us to apply HQET machinery for
the evaluation of coefficients in the expansion of $W_C$:
\be
W_C = 1+ a_1 (iy_-v_+) + a_2 (iy_-v_+)^2 + a_3 (iy_-v_+)^3 + ...\ ,
\lab{smal}
\ee
where
$$
a_1 = v_+^{-1} \bra{B} \Phi_v^\dagger (i D_+) \Phi_v \ket{B}=0
\,,\quad
a_2 = -\frac16 \bra{B} \Phi_v^\dagger (i D)^2 \Phi_v \ket{B}
\,,\quad
a_3 = -\frac1{18}\bra{B} \Phi_v^\dagger v_\mu F_{\mu\nu} D_\nu \Phi_v \ket{B}
{}.
$$
We recognize that our ansatz \re{ans} is consistent at $\CO(y_-^2)$
with the small $y_-$ expansion of $W_C$. Note that in the leading $1/M$
limit, the heavy quark fields $h_v$ are proportional to the Wilson line
$\Phi_v$, which is why the coefficients $a_n$ are equal to analogous
fundamental parameters in HQET \ci{Bigi,Neub}.
In particular, we can identify the correlation length $l$ entering into
\re{ans} as
\be
l^2 = -a_2 = -6/\lambda_1 = 6/\mu_\pi^2\,.
\lab{A}
\ee
where the value of $\mu_\pi^2=0.52\pm 0.12 (\GeV)^2$
was estimated using QCD sum rules in \ci{Sum}
(which implies a particular choice for the perturbation series).
Substituting the nonperturbative estimate \re{ans}
for the Wilson line $W_C$ into
\re{f}, we find a distribution function $f^{(0)}(x)$, which can be
used as a nonperturbative input for the evolution equations \re{eq1}. If we
use only the small
$y_-$ expansion \re{smal} we get an expression for $f^{(0)}(x)$ as a
series of derivatives of $\delta(1-x)-$function, equivalent to that
proposed in \ci{Bigi,Neub}, which requires us to guess
the shape
of the function in the end-point region $x\sim 1$.
On the other hand,
integrating the nonperturbative ansatz \re{ans}, inspired by the IR renormalon
analysis, we find the function
\footnote{The same function has been proposed in \ci{Neub}.}
\be
f^{(0)}(x)=\frac1{\sqrt{2\pi\sigma^2}}
\exp\lr{-\frac{(1-x)^2}{2\sigma^2}}\,,\qqquad
\sigma^2=\frac2{l^2M^2}=\frac{\mu_\pi^2}{3M^2}\,,
\lab{f0}
\ee
which describes a Gaussian distribution around $x=1$ with the
width $\sigma$. Notice however, that
since we neglected $\CO(y_-^3)$ corrections to the exponent of $W_C$
in \re{ans},
this function does not vanish outside the physical region $x>M_B/M$ and
\re{f0} is valid only at the vicinity of $x=1$.
To get a better approximation for the distribution function one needs
a more realistic estimate for the Wilson line expectation value.
Once we know the distribution function $f^{(0)}(x)$, we evaluate
its moments, $f^{(0)}_n=\int_0^{\xm}dx\ x^{n-1}f^{(0)}(x)$, and
substitute them into \re{soln} and \re{res1}.

\section{Factorization in semileptonic decay of the B meson}

The factorization for $\B\to \gamma X_s$ was based on the kinematical
requirement that, near the end point of the photon spectrum, the $\s$ quark
must form a jet with large energy and small invariant mass.
The same situation appears in semileptonic decay $\B\to l\barnu X_u$
in the end-point region of the charged lepton spectrum.
In this case, the decay occurs through an effective local vertex
$\bar u\gamma_\mu(1-\gamma_5)b\cdot \bar l \gamma_\mu(1-\gamma_5)\nu$.
Let $l$ and $\nu$ be the momenta of the leptons, and $Q$ and $P$ the
momenta of the $\b$ and $\u$ quarks, respectively.
Standard notations for the scaling
variables in the rest frame of the $\B$ meson are,
\be
x=\frac{2E_l}{M}\,,\qqqquad
y=\frac{W^2}{M^2}\,,\qqqquad
y_0=\frac{2W_0}{M}\,,
\lab{scal}
\ee
where $W=l+\nu=Q-q$ is the momentum transferred to the leptons.
Phase space for these variables is
\be
0 \leq x \leq x_m=\xm,\qqquad
0 \leq y \leq x x_m,\qqquad
y/x + x \leq y_0 \leq y/x_m + x_m\,.
\lab{ps}
\ee
In analogy to the radiative decay, we examine the energy and invariant mass
of the jet created by the $\u$ quark,
\be
P^2=M^2 (1-y_0+y), \qquad
P_0=\frac{M}2(2-y_0)\,.
\ee
{}From the definition \re{ps} of phase space we find that
\be
P^2\leq M^2(1-x)(1-y/x),\qquad
P_0 \leq \frac{M}2(1-y) +\frac{M}2(1-x)(1-y/x)\,.
\lab{cond}
\ee
We conclude that near the end point of the lepton spectrum, $x\to 1$,
with $y < 1$,
we meet the same kinematical situation as in the $\B\to\gamma X_s$ decay.
Thus, once again,
in the leading $1/M$ limit the dominant contribution to the
differential rate of the decay comes from diagrams containing hard,
jet and soft subprocesses. As in the previous case, the contributions of
these subprocesses are factorized from the partonic
subprocess $\sigma_0\sim(Q\cdot\nu)(l \cdot P)=\frac{M^4}4(x-y)(y_0-x)$,
and for the triple differential rate we get an expression similar to \re{fac},
\be
\frac1{\Gamma_l}\frac{d^3 \Gamma}{dx dy dy_0}=
(x-y)(y_0-x)\ \frac{M}{v_+}\int_{(l_+)_{min}}^{P_+} dl_+\ S(l_+)
J(P_+-l_+)H(P_-)\,.
\lab{fac1}
\ee
where $\Gamma_l=\frac{G_F^2}{16\pi^3} |V_{ub}|^2 M^5$, and $(l_+)_{min}$
was defined in \re{lmin}. Note that the prefactor in \re{fac1}
suppresses the ``dangerous'' region $y\to x\sim 1$.
In this region, $P_0$ is forced to zero by \re{cond} and
factorization fails because the outgoing quark has vanishing energy.
We note that relations \re{fac}
and \re{fac1} were found in a frame,
where the outgoing quark has momentum
$P_+ \ll P_-$. For the radiative
decay, this frame is fixed by the particular choice \re{q}
of the photon momentum $q$.
For a semileptonic decay, we use the same choice for the lepton momentum,
$l=\frac{M}{\sqrt 2}(x,0,\vec 0)$. Let us express $P_+$ and $P_-$ in terms
of the scaling variables \re{scal}. Using the relation $P_+ \ll P_-$ we
make the following approximation:
$P_-\approx P_-+P_+=\sqrt{2}P_0=\sqrt{2}(M-W_0)$
and get
\be
P_-=\frac{M}{\sqrt 2}(2-y_0)\,,\qquad
P_+=\frac{P^2}{2P_-}=\frac{M}{\sqrt 2}(1-\xi)\,,\qquad
\xi=\frac{1-y}{2-y_0}\,,
\lab{kin}
\ee
where the scaling variable $\xi$ is an analog of $x$ in the radiative
decay, eq.\re{fac}. However, in contrast to the previous case, $\xi$ depends on
$y$ and $y_0$, and to calculate  the differential rate
${d\Gamma_{\B\to l\barnu X_u}}/{dx}$ one has to integrate \re{fac1}
with respect to
$\xi$ in the physical region $x\leq \xi \leq M_B/M$.
Using the kinematic relations \re{kin} and the definitions \re{f} and \re{J}
of the subprocesses we can rewrite \re{fac1}
as
\be
\frac1{\Gamma_l}\frac{d^3 \Gamma}{dx dy dy_0}=(x-y)(y_0-x)
 \int_\xi^{\xm} dz\ f\lr{z,\frac{M}{\mu}}
M^2 J(M^2(2-y_0)(z-\xi),\mu^2) H\fr{M(2-y_0)}{\mu}\,.
\lab{fac2}
\ee
Notice that if we replace subprocesses
by their lowest order expressions $H=1$, $f=f^{(0)}(z)$,
$J=\delta\lr{M^2(2-y_0)(z-\xi)}$, we get a result which contains all
nonperturbative corrections, and which coincides with
analogous expression in \ci{Bigi}.

In order to analyze large perturbative
corrections to $d\Gamma/dx$ in the end-point region, we change variables
from $y$ to $\xi$, and integrate \re{fac2} with respect to $\xi$ and $y_0$,
neglecting terms that vanish as $x\to 1$,
\ba
\frac1{\Gamma_l}\frac{d\Gamma}{dx}=\int_1^2 dy_0\ (2-y_0)^2 (y_0-1)
\int_x^{\xm} d \xi \int_\xi^{\xm} dz\ &&
\lab{new1}
\\
&&\hspace{-75mm}
\times
f\lr{z,\frac{M}{\mu}}\
M^2 J\lr{M^2(2-y_0)(z-\xi),\mu^2}\ H\fr{M(2-y_0)}{\mu}
+\CO(1-x)\,.
\nonumber
\ea
Comparing this expression with \re{new} we notice that
$-\frac{d\ }{dx}\lr{\frac1{\Gamma_l}\frac{d\Gamma}{dx}}$
is similar, apart from $y_0-$integral, to the differential
rate $\frac1{\Gamma_\gamma}\frac{d\Gamma}{dx}$ of the radiative
decay, eq.\re{new}. This suggests the appropriate moments
for semileptonic decay
in analogy with \re{Mn},
\be
{\cal M}_n\lr{\B\to l\barnu X_u}\equiv
-\int_0^{\xm} dx x^{n}\frac{d\ }{dx}\lr{\frac1{\Gamma_l}\frac{d\Gamma}{dx}}
=\frac{n}{\Gamma_l}\int_0^{\xm} dx x^{n-1}\frac{d\Gamma}{dx}\,.
\lab{Mn1}
\ee
As in the case of radiative decays, we evaluate the moments of
$\frac{d\Gamma}{dx}$, eq.\re{new1}, and find in the large $n$ limit
the following relation
\be
{\cal M}_n\lr{\B\to l\barnu X_u}=f_n\fr{M}{\mu}
\int_0^1 dx_\nu\ x_\nu (1-x_\nu)\
J_n\fr{M\sqrt{1-x_\nu}}{\mu}H\fr{M(1-x_\nu)}{\mu} + \CO(1/n)
\lab{res2}
\ee
where we have changed variables to
$y_0=1+x_\nu\approx 1+\frac{2E_\nu}{M}$ with $E_\nu$
the energy of outgoing neutrino in the rest frame of the $\B$ meson.
Here, the moments $f_n$ and $J_n$ are {\it identical\/} to whose in
the radiative decay and satisfy the evolution equations \re{eq1} and
\re{eq2}. The only difference with \re{res} is that one has to integrate
in \re{res2} with respect to the energy of the neutrino, $x_\nu$, and
take into account the $x_\nu-$dependence of the collinear and hard
subprocesses. This additional integration
does not affect leading double logarithmic corrections $(\as\log^2n)^k$,
but it does affect nonleading terms. Note that in \re{res2}
the prefactor suppresses the end-points $x_\nu=0$ and
$x_\nu=1$, which correspond to the limits of soft neutrino and outgoing quark,
respectively. Thus, our factorized expression \re{res2} takes care by
itself near the ``dangerous'' point $x_\nu=1$, where IR factorization
fails.

Let us compare our predictions \re{res} and \re{res2} for resummed
large perturbative corrections with the results of one-loop calculations
\ci{Ali,Jezabek}
of the differential rate for $\B\to \gamma X_s$ and $\B\to l\barnu X_u$
in the end-point region. Solving the evolution equations
\re{eq1} and \re{eq2} we find one-loop expressions for the subprocesses:
\ba
f_n(M/\mu)&=&1+\alpi C_F\left(
              -\log^2\frac{\mu n}{Mn_0} + \log\frac{\mu n}{Mn_0}\right)
\nonumber
\\
J_n(M/\mu)&=&1+\alpi C_F\left( 2\log^2\frac{\mu \sqrt n}{M\sqrt{n_0}}
              +\frac32 \log\frac{\mu\sqrt n}{M\sqrt{n_0}}\right)
\\
H(M/\mu)&=&1+\alpi C_F\left(-\log^2\frac{\mu}{M} - \frac52 \log\frac{\mu}{M}
\right)
\nonumber
\ea
where we have omitted constant terms and $f_n^{(0)}$.
Substituting these relations into \re{res} and \re{res2}, we obtain
the one-loop expression for the moments of the differential rates
\be
{\cal M}_n \sim
1+\frac{\as}{2\pi} C_F\lr{-\log^2 n + A\ \log n + \mbox{const.}}
\ee
where coefficients in front of nonleading $\log n-$term are
\be
A_{\B\to \gamma X_s}=\frac72\,,\qqquad A_{\B\to l\barnu X_u}=\frac{31}6
\ee
in accordance with the one-loop results of ref.\ci{Ali}
and \ci{Jezabek}.

Using the evolution equations \re{eq1} and \re{eq2}, one can represent
the expression \re{res2}
in a form similar to \re{res1}. It is more interesting however, to consider
the ratio of the moments ${\cal M}_n\lr{\B\to\gamma X_s}$ and
${\cal M}_n\lr{\B\to l\barnu X_u}$ defined in \re{res} and \re{res2},
respectively. We find that the moments of the heavy quark distribution
function, $f_n$, cancel in the ratio and we get
\ba
\frac{{\cal M}_n\lr{\B\to l\barnu X_u}}{{\cal M}_n\lr{\B\to\gamma X_s}}
&=&\frac{C_l}{C_\gamma}
\int_0^1 dx_\nu\ x_\nu(1-x_\nu)
\exp\lr{2 \int_{1-x_\nu}^1\frac{dy}{y}\int_{M\sqrt{\frac{yn_0}{n}}}^{My}
        \frac{dk_t}{k_t}\Gamma_{cusp}(\as(k_t))}
\lab{ratio}
\\
&\times&
\exp\lr{
\int_{1-x_\nu}^1\frac{dy}{y}
\left[-\gamma\lr{\as\lr{M\sqrt{{yn_0}/{n}}}}
      +2\gamma(\as(My))+\Gamma(\as(My))
\right]
}
\,.
\nonumber
\ea
where $C_l=1+\CO(\as(M))$ is an analog of $C_\gamma$, defined in \re{res1},
for the case of semileptonic decay.
Thus, all nonperturbative corrections cancel in the ratio of the moments of
the differential rates and this allows us to calculate \re{ratio}
perturbatively.
We may then compare it with the ratio of experimental data using the
definitions of the moments, \re{Mn} and \re{Mn1}, and isolating the ratios
of the elements of the Cabibbo-Kobayashi-Maskawa matrix contained
in the prefactors $\Gamma_\gamma$ and $\Gamma_l$.

\section{Conclusions}

In this paper we performed infrared factorization on the differential
rates of radiative and semileptonic inclusive
decays of the $\B$ meson in the end-point
regions of the photon and the charged lepton spectra. We found that in the
leading $1/M$ limit the differential rates are expressed in terms of
hard $(H)$, jet $(J)$ and soft $(S)$ functions which satisfy evolution
equations.

Solving the evolution equations we found expressions for the moments
of the differential rates in the end-point region, \re{res1} and \re{ratio},
which take into
account all leading and nonleading logarithmic $\log n$
corrections in perturbation theory, as well as
large nonperturbative power corrections
in the leading $1/M$ limit. Expanding these expressions in
powers of the coupling constant and using one-loop results \re{one}
for the anomalous
dimensions entering the evolution equations, we have shown that
our predictions coincide to the lowest order with the results of
one-loop calculations for both processes.

Nonperturbative corrections appear in our formalism from the boundary
value of the soft function $f^{(0)}(x)$ in the evolution equation \re{eq1}.
The soft function is the universal process-independent function which
describes the distribution of the $\b$ quark in the $\B$ meson.
We established that the behavior of $f^{(0)}(x)$ in the end-point region
$x\sim 1$ is governed by the nonperturbative asymptotics of the
Wilson line expectation values. Considering the contribution of
infrared renormalons, we found a nonperturbative
ansatz for the Wilson line which led to a Gaussian model \re{f0} for the
heavy quark distribution function $f^{(0)}$.

\medskip


\bigskip\par\noindent{\Large {\bf Acknowledgements}}\par\bigskip\par

\noindent
We are grateful for helpful conversations with R.Akhoury, A.Radyushkin
and M.Wise. One of us (G.P.K.) would like to thank M.Fontannaz, M.Neubert
and D.Schiff for useful discussions and hospitality at LPTHE (Orsay) and
CERN. This work was supported in part by the National Science Foundation
under grant PHY9309888.

\bb{99}
\bi{review}
     A.Le Yaouanc, L.Oliver, O.Pene and J.C.Raynal,
     Talk given at 29th Rencontres de Moriond: QCD and High Energy Hadronic
     Interactions, Meribel les Allues, France, 19-26 Mar 1994;
     hep-ph/9406341.
\bi{Bigi}
     I.I.Bigi, M.A.Shifman, N.G.Uraltsev and A.I.Vainstein,
     preprint CERN-TH.7129/93, Dec. 1993; hep-ph/9312359;
\\   M.Shifman, preprint TPI-MINN-94-17-T, Feb 1994; hep-ph/9405246.
\bi{Neub}
     M.Neubert,  Phys. Rev. D49 (1994) 4623,\ 3392.
\bi{ACM}
     G.Altarelli, N.Cabibbo, G.Corbo, L.Maiani and G.Martinelli,
     Nucl. Phys. B208 (1982) 365.
\bi{Wise}
     A.F.Falk, E.Jenkins, A.V.Manohar and M.B.Wise,
     Phys. Rev. D49 (1994) 4553.
\bi{theorems}
     J.C.Collins, D.E.Soper and G.Sterman, ``Factorization of Hard Processes
     in QCD'',
     in {\it Perturbative Quantum Chromodynamics\/}, ed. by A.H.Mueller (World
     Scientific, Singapore, 1989) p.1.
\bi{IRR1}
     A.H.Mueller, Nucl. Phys. B250 (1985) 327;
     in {\it QCD 20 years later}, Aachen, 1992,
     Eds. P.M.Zerwas and H.A.Kastrup,
     World Scientific, Singapore, 1993.
\\   M.Beneke and V.I.Zakharov, Phys. Lett. B312 (1993) 340;
\\   V.I.Zakharov, Nucl. Phys. B385 (1992) 452.
\bi{IRR2}
     M.Beneke and V.M.Braun, preprint MPI-PHT-94-9, Feb 1994;
     hep-ph/9402364;
\\   A.V.Manohar and M.B.Wise, preprint UCSD-PTH-94-11, Jun 1994;
     hep-ph/9406392.
\bi{IRR3}
     H.Contopanagos and G.Sterman, Nucl. Phys. B419 (1994) 77.
\bi{tech}
     G.Sterman, Nucl. Phys. B281 (1987) 310;
\\   S.Catani and L.Trentadue, Nucl. Phys. B327 (1989) 323; B353 (1991) 183.
\\   G.P.Korchemsky and G.Marchesini, Phys. Lett. B313 (1993) 433
\bi{S}
     G.P.Korchemsky, Mod. Phys. Lett. A4 (1989) 1257.
\bi{YaF}
     S.V.Ivanov, G.P.Korchemsky and A.V.Radyushkin, Sov. J. Nucl. Phys.
     44 (1986) 145;
\\   G.P.Korchemsky and A.V.Radyushkin, Sov. J. Nucl. Phys. 45 (1987) 127, 910.
\bi{GLAP}
     V.N.Gribov and L.N.Lipatov, Yad. Fiz. 15 (1972) 781; 1218;
\\   L.N.Lipatov, Yad. Fiz. 20 (1974) 181;
\\   G.Altarelli and G.Parisi, Nucl. Phys. 126B (1977) 298.
\bi{Effe}
     M.A.Shifman, A.I.Vainshtein and V.I.Zakharov,
     Phys. Rev. D18 (1978) 2583
\bi{Jezabek}
     G.Corbo, Nucl. Phys. B212 (1983) 99;
\\   M.Je\.zabek and J.H.K\"uhn, Nucl. Phys. B320 (1989) 20.
\bi{eik=WL}
     M.Ciafaloni and G.Curci, Phys. Lett. B102 (1981) 352;
\\   J.C.Collins and D.E.Soper, Nucl. Phys. B193 (1981) 381.
\bi{eik}
     A.Bassetto, M.Ciafaloni and G.Marchesini, Phys. Rep. 100 (1983) 201.
\bi{Wn}
     G.P.Korchemsky and G.Marchesini, Nucl. Phys. B406 (1993) 225.
\bi{KR}
     G.P.Korchemsky and  A.V.Radyushkin, Phys. Lett. B279 (1992) 359.
\bi{ren}
     A.M.Polyakov, Nucl. Phys. B164 (1980) 171;
\\   I.Ya.Aref'eva, Phys. Lett. B93 (1980) 347;
\\   V.S.Dotsenko and S.N.Vergeles, Nucl. Phys. B169 (1980) 527;
\\   R.A.Brandt, F.Neri and M.-A.Sato, Phys. Rev. D24 (1981) 879.
\bi{2-loop}
     G.P.Korchemsky and A.V.Radyushkin,
     Phys. Lett. B171 (1986) 459; Nucl. Phys. B283 (1987) 342.
\bi{WL}
     I.A.Korchemskaya and G.P.Korchemsky, Phys. Lett. B287 (1992) 169;
\\   A.Bassetto, I.A.Korchemskaya, G.P.Korchemsky and G.Nardelli,
     Nucl. Phys. B408 (1993) 62.
\bi{KS} G.P.Korchemsky and G.Sterman, in preparation.
\bi{Gath}
     G.Sterman, in {\it Perturbative Quantum Chromodynamics\/},
     (Tallahassee, 1981) Ed. D.W.Duke and J.F.Owens,
     AIP Conference Proceedings, No. 74, NY 1981, p.22.
\\   J.G.M.Gatheral, Phys. Lett. 133B (1984) 90;
\\   J.Frenkel and J.C.Taylor, Nucl. Phys. B246 (1984) 231.
\bi{kt}
     D.Amati, A.Bassetto, M.Ciafaloni, G.Marchesini and G.Veneziano,
     Nucl. Phys. B173 (1980) 429.
\bi{Geor}
     H.Georgi, ``Heavy Quark Effective Field Theory'', in
     {\it Boulder TASI Proceedings 91}, p.589 (QCD161:T45:1991).
\bi{Sum}
     P.Ball and V.M.Braun, Phys. Rev. D49 (1994) 2472.
\bi{Ali}
     A.Ali and C.Greub, Phys. Lett. B287 (1992) 191.
\eb

\newpage
\begin{center}
\unitlength=1.0mm
\linethickness{0.4pt}
\begin{picture}(107.00,103.00)(0,60)
\put(30.00,130.00){\circle{14.00}}
\put(85.00,130.00){\circle{14.00}}
\put(85.00,130.00){\makebox(0,0)[cc]{$H$}}
\put(30.00,130.00){\makebox(0,0)[cc]{$H$}}
\put(58.00,130.00){\oval(18.00,14.00)[]}
\put(58.00,130.00){\makebox(0,0)[cc]{$J$}}
\put(37.00,130.00){\line(1,0){12.00}}
\put(67.00,130.00){\line(1,0){11.00}}
\put(20.00,84.00){\line(1,5){7.86}}
\put(95.00,84.00){\line(-1,5){7.86}}
\put(58.00,100.50){\oval(28.00,19.00)[]}
\put(58.00,101.00){\makebox(0,0)[cc]{$S$}}
\put(16.00,96.00){\makebox(0,0)[cc]{$Q$}}
\put(100.00,96.00){\makebox(0,0)[cc]{$Q$}}
\put(58.00,116.00){\makebox(0,0)[cc]{$\ldots$}}
\put(86.00,97.00){\makebox(0,0)[cc]{$\vdots$}}
\put(30.00,102.00){\makebox(0,0)[cc]{$\vdots$}}
\put(8.00,139.00){\makebox(0,0)[cc]{$q$}}
\linethickness{0.4pt}
\put(91.50,136.50){\oval(5.00,5.00)[rb]}
\put(96.50,136.50){\oval(5.00,5.00)[lt]}
\put(96.50,141.50){\oval(5.00,5.00)[rb]}
\put(101.50,141.50){\oval(5.00,5.00)[lt]}
\put(23.50,136.50){\oval(5.00,5.00)[lb]}
\put(18.50,136.50){\oval(5.00,5.00)[rt]}
\put(18.50,141.50){\oval(5.00,5.00)[lb]}
\put(13.50,141.50){\oval(5.00,5.00)[rt]}
\put(107.00,139.00){\makebox(0,0)[cc]{$q$}}
\linethickness{1.5pt}
\bezier{5}(53.00,110.00)(53.00,117.00)(53.00,123.00)
\bezier{5}(62.00,123.00)(62.00,117.00)(62.00,110.00)
\bezier{8}(70.00,108.00)(80.00,111.00)(89.00,113.00)
\bezier{7}(72.00,101.00)(82.00,102.00)(91.00,103.00)
\bezier{8}(71.00,94.00)(82.00,92.00)(93.00,92.00)
\bezier{7}(26.00,109.00)(35.00,107.00)(44.00,105.00)
\bezier{8}(23.00,97.00)(34.00,97.00)(44.00,97.00)
\linethickness{0.4pt}
\put(19.00,88.00){\line(1,2){3.00}}
\put(22.00,94.00){\line(1,-6){1.11}}
\put(94.00,89.00){\line(1,6){1.01}}
\put(94.00,89.00){\line(-1,2){3.00}}
\put(13.00,144.00){\line(1,-2){2.0}}
\put(13.00,144.00){\line(1,0){3.50}}
\put(100.00,143.00){\line(1,2){2.0}}
\put(100.00,143.00){\line(1,0){3.50}}
\put(61.50,149.50){\oval(7.00,7.00)[lt]}
\put(56.00,73.00){\oval(8.00,6.00)[rb]}
\put(58.00,144.00){\line(0,-1){5.00}}
\put(60.00,76.00){\line(0,1){7.00}}
\put(60.00,87.00){\line(0,1){3.00}}
\end{picture}
\par\parbox[t]{140mm}
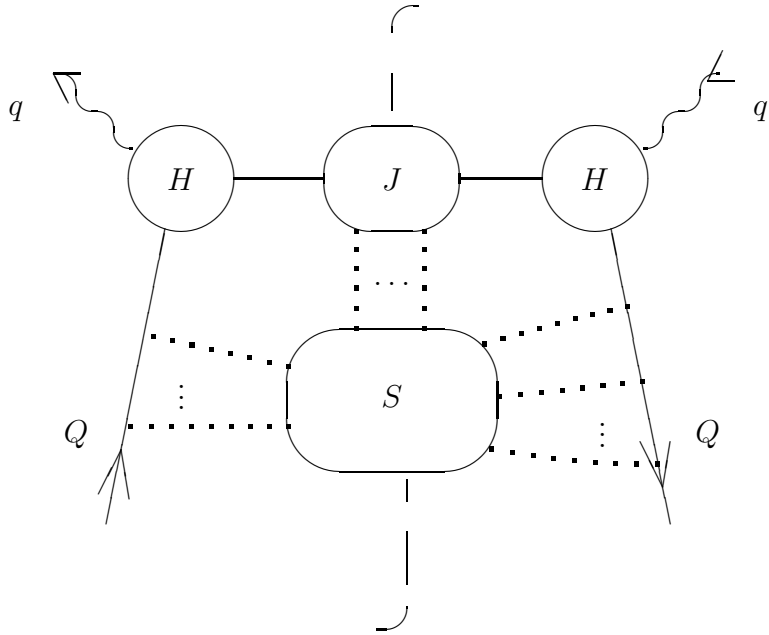
{Fig.1: Unitarity diagram contributing to the differential rate of
the radiative inclusive $\B\to\gamma X_s$ decay in the leading $1/M$
limit in the axial gauge. We use solid lines for quarks, dotted lines
for gluons, wave lines for photons and a dashed line for the final state.}

\unitlength=0.7mm
\linethickness{0.4pt}
\begin{picture}(83.00,116.00)
\put(20.00,20.00){\line(0,1){70.00}}
\put(20.00,90.00){\line(1,0){55.00}}
\put(75.00,90.00){\line(0,-1){70.00}}
\put(22.00,20.00){\line(0,1){68.00}}
\put(22.00,88.00){\line(1,0){51.00}}
\put(73.00,88.00){\line(0,-1){68.00}}
\put(15.00,96.00){\makebox(0,0)[cc]{$0$}}
\put(72.00,96.00){\makebox(0,0)[cc]{$y=(0_+,y_-,\vec 0)$}}
\put(13.00,39.00){\makebox(0,0)[cc]{$v$}}
\put(83.00,39.00){\makebox(0,0)[cc]{$-v$}}
\put(21.00,14.00){\makebox(0,0)[cc]{$\vdots$}}
\put(74.00,14.00){\makebox(0,0)[cc]{$\vdots$}}
\put(21.00,42.00){\line(-2,-5){4.03}}
\put(21.00,42.00){\line(2,-5){3.89}}
\put(74.00,32.00){\line(-2,5){3.95}}
\put(74.00,32.00){\line(2,5){3.97}}
\put(49.00,89.00){\line(-5,2){7.06}}
\put(49.00,89.00){\line(-5,-2){7.06}}
\end{picture}
\par
\parbox[t]{140mm}
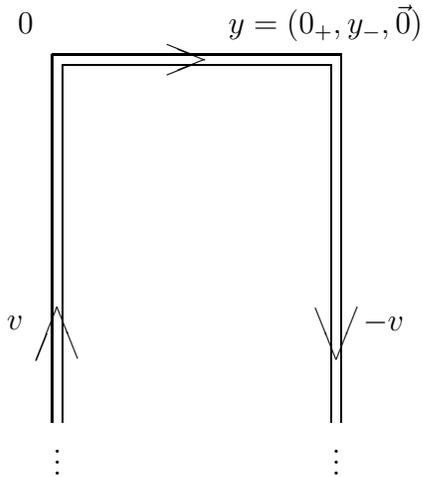
{Fig.2: Integration path $C$ entering into the definition \re{S} of the
Wilson line $W_C$.}

\end{center}

\end{document}